\documentclass[journal]{IEEEtran}

\usepackage{amsmath,amssymb,amsfonts}
\usepackage{graphicx}
\graphicspath{{./}{arxiv/}}
\usepackage{textcomp}
\usepackage{xcolor}
\usepackage{booktabs}
\usepackage{multirow}
\usepackage{url}
\usepackage{cite}
\usepackage{algorithm}
\usepackage{algorithmic}
\usepackage{subcaption}
\usepackage{balance}
\usepackage{float}

\newcommand{\best}[1]{\setbox0\hbox{#1}\rlap{\rule[-2.4pt]{\wd0}{0.8pt}}\box0}
\begin{document}

\title{Frequency-Ordered Tokenization for Better Text Compression}

\author{Maximilian Kalcher, ETH Zurich (mkalcher@ethz.ch)}

\maketitle

\begin{abstract}
We present frequency-ordered tokenization (FOT), a simple preprocessing technique that improves lossless text compression by exploiting the power-law frequency distribution of natural language (Zipf's law). FOT segments text into the vocabulary of a pretrained Byte Pair Encoding (BPE) tokenizer, renumbers tokens by frequency with a fixed table, and writes them with a short byte code before any standard compressor. Because the FOT stream is less than half the size of the text, compressors produce smaller output and run faster at the same time. On enwik8 (100\,MB of Wikipedia), FOT reduces the compressed size by 7.96 percentage points (pp) for zlib, 2.24\,pp for xz (LZMA), 1.96\,pp for brotli and 1.44\,pp for zstd, while making total compression, including the transform, $1.6\times$, $3.2\times$, $1.8\times$ and $5.0\times$ faster. Gains hold from 100\,KB to 1\,GB (enwik9) and across texts of diverse languages, and outperform the classical Word Replacing Transform. On enwik9, FOT with tuned xz sets a new \emph{state-of-the-art} among LZ-based compressors on the Large Text Compression Benchmark~\cite{mahoney2006}.
\end{abstract}

\begin{IEEEkeywords}
Lossless compression, text compression, Zipf's law, Byte Pair Encoding, tokenization, preprocessing
\end{IEEEkeywords}

\section{Introduction}

Lossless text compression is fundamental to modern computing. Shannon~\cite{shannon1948} established that the entropy of a source is the theoretical lower bound on compression, and that better statistical models of the source enable better compression.

Practical compressors are either \emph{dictionary-based} (LZ77~\cite{ziv1977} and its descendants zlib~\cite{gailly1995}, zstd~\cite{collet2016}, brotli~\cite{alakuijala2018} and LZMA~\cite{pavlov1998}), exploiting repeated byte patterns, or \emph{statistical} (PPM~\cite{cleary1984}), modeling symbol probabilities in context. Neural compressors such as cmix~\cite{knoll2024} and NNCP~\cite{bellard2021} reach ratios below 15\% on standard benchmarks, but at 100--1000$\times$ the computational cost of practical compressors.

We propose \emph{frequency-ordered tokenization} (FOT), a preprocessing step that transforms text into a representation more amenable to LZ-family compression. Natural-language token frequencies follow Zipf's law~\cite{zipf1949}: a small fraction of tokens accounts for most occurrences. While prior word-level transforms~\cite{kruse1998,skibinski2005} exploit similar intuitions, we operate at the \emph{subword} level and reuse the vocabularies that large language models already train at great expense~\cite{radford2019,tiktoken2023}. By (1) segmenting text into such a vocabulary, (2) assigning small integer IDs to frequent tokens, and (3) encoding with a variable-length byte code, we produce fewer, more repetitive bytes that LZ compressors exploit efficiently. Unlike word-level methods, BPE handles morphological variation, out-of-vocabulary tokens, and non-Latin scripts without language-specific rules.

\textbf{Contributions.} (1) FOT improves LZ-family compression by 1.3--8\,pp on enwik8/enwik9 with negligible per-file overhead, outperforming the classical Word Replacing Transform~\cite{skibinski2005}. (2) It also accelerates compressors (up to $5.0\times$ for zstd-22), a Pareto improvement on the speed--ratio tradeoff, and a minimum-code-length segmentation into the pretrained vocabulary beats exact BPE inference. (3) We give ablations, per-file overhead and information-theoretic analyses, and results on Chinese and Arabic text.

\section{Background and Related Work}

\subsection{Zipf's Law and Variable-Length Encoding}

\begin{figure}[H]
\centering
\includegraphics[width=\columnwidth]{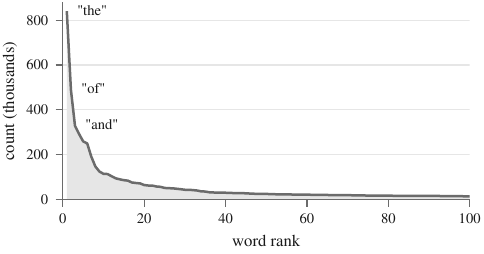}
\caption{Zipf's law on enwik8: frequency of the 100 most common (lowercased) words vs.\ rank. The top 10 words account for over 20\% of all occurrences.}
\label{fig:zipf_bg}
\end{figure}

Zipf's law~\cite{zipf1935,zipf1949} states that the frequency of the $r$-th ranked word in a corpus is $f(r) \propto r^{-\alpha}$, with $\alpha \approx 1$ for natural language~\cite{piantadosi2014}. This concentration (Fig.~\ref{fig:zipf_bg}) motivates short codes for frequent tokens.

We encode token ranks with a byte-oriented prefix code: ranks 0--127 take 1 byte, the next $126 \cdot 256 = 32{,}256$ ranks take 2 bytes (lead byte 128--253) and the remaining ranks take 3 bytes (lead byte 254--255), most significant part first. Like the universal codes~\cite{elias1975}, it suits distributions concentrated on small integers, and it produces ordinary bytes for any compressor. Compared to the LEB128 varint~\cite{google2008} (7 bits per byte, least significant first) its 2-byte band is twice as wide; on enwik8 this shortens the stream by 3\% and improves compression by 0.1--0.4\,pp for zlib, xz and zstd (Table~\ref{tab:ablation}).

\subsection{Byte Pair Encoding and Compressors}

BPE~\cite{gage1994,sennrich2016} iteratively merges frequent symbol pairs to build a subword vocabulary and is the dominant tokenizer for large language models~\cite{radford2019,brown2020}. Byte-level BPE includes every single byte in the vocabulary, so \emph{any} segmentation of a byte sequence into vocabulary tokens is lossless, not only the one BPE's merge rules produce. LZ77-family compressors (zlib~\cite{gailly1995}, zstd~\cite{collet2016}, brotli~\cite{alakuijala2018}, LZMA~\cite{pavlov1998}) code literals with little or no context (order-0 Huffman in zlib and zstd, low-order modeling in brotli and LZMA), leaving room for preprocessing; statistical compressors (PPMd~\cite{shkarin2002}, cmix~\cite{knoll2024}) already capture frequency structure.

\subsection{Related Work}

Preprocessing data before compression has a long history. The BWT~\cite{burrows1994} groups similar contexts; move-to-front coding assigns small integers to recent symbols, a conceptual precursor to our frequency ordering. At the word level, Kruse and Mukherjee~\cite{kruse1998} replaced words with dictionary codes (up to 20\% gains for bzip2), and Skibi\'{n}ski et al.~\cite{skibinski2005} introduced the Word Replacing Transform (WRT), which replaces words with frequency-ranked codes and improves on the earlier StarNT preprocessor by 2--6\% for PPM and BWT compressors; an LZ77-oriented variant reaches up to 14\% with gzip. Brotli~\cite{alakuijala2018} ships a static 120\,KB dictionary of common words, which is why it compresses small text files well.

Grammar-based compressors such as GLZA~\cite{conrad2016} build a phrase grammar for each file and code it with context models, reaching PPM-class ratios at a much higher compression cost; FOT instead reuses a fixed pretrained vocabulary to make standard LZ compressors smaller and faster. Del\'{e}tang et al.~\cite{deletang2024} found that all but the smallest \emph{neural} compressors compress better with simpler tokenizers, as they already capture the patterns tokenization exposes. Our work targets classical LZ compressors, where the benefit is complementary. Concurrent with our work, Erdogan et al.~\cite{erdogan2026} study BPE before LZ compression, reporting 10--20\% improvements; our contribution adds frequency-ordered ID assignment with a static rank table, the code-length-optimal segmentation, and the speed--ratio Pareto improvement.

\section{Method}

\subsection{Frequency-Ordered Tokenization}

Given input text $T$, a byte-level BPE vocabulary $\mathcal{V}$, and a rank table $\pi$ (a bijection from token IDs to ranks, frequent tokens first):

\begin{algorithm}[H]
\caption{Frequency-Ordered Tokenization}
\begin{algorithmic}[1]
\REQUIRE Text $T$, vocabulary $\mathcal{V}$, static rank table $\pi_s$
\STATE $\mathbf{t} \leftarrow \text{Segment}(T, \mathcal{V})$ \hfill $\triangleright$ Token sequence
\STATE $\pi \leftarrow \pi_s$, optionally prefixed by the $K$ most frequent tokens of $\mathbf{t}$ \hfill $\triangleright$ Rank map
\STATE $t'_j \leftarrow \pi(t_j)$ for all $j$ \hfill $\triangleright$ Reorder IDs
\STATE $B \leftarrow \text{PrefixVarint}(\mathbf{t}')$ \hfill $\triangleright$ Byte stream
\RETURN $\text{Compress}(B)$ \hfill $\triangleright$ Any compressor
\end{algorithmic}
\end{algorithm}

Decompression inverts each step and concatenates the token byte strings; the pipeline is $O(N)$.

\textbf{Rank table.} The simplest choice computes $\pi$ from the frequencies of the file itself and stores it alongside the compressed data (\emph{adaptive}). This table costs about $2.3U$ bytes for $U$ distinct tokens, which is negligible at 100\,MB but 5--11\% of the file size below 1\,MB (Section~\ref{sec:overhead}). We therefore use a \emph{static} table $\pi_s$, computed once from a disjoint corpus and shipped with the vocabulary (400\,KB for 100k tokens), so that no full table is stored per file. The encoder may additionally prefix $\pi_s$ with an explicit list of the file's $K$ most frequent tokens (\emph{hybrid}: 128 tokens or 0.3\,KB on enwik8; with $K = 0$ the transform is a single pass); since only the band boundaries of the prefix code affect the stream length, a short list already captures most of the adaptive benefit. $K \in \{0, 32, 64, \dots\}$ is chosen by minimizing $0.1\,S_K + 0.85\,M_K$, where $S_K$ is the varint stream length and $M_K$ the list size, coefficients fitted once on zstd, xz, zlib and brotli: compressors realize only about a tenth of the raw bytes saved by reordering, while the ID list barely compresses.

\textbf{Segmentation.} The reference segmentation is exact BPE inference (tiktoken~\cite{tiktoken2023}). Since every segmentation into $\mathcal{V}$ round-trips, we also use a \emph{minimum-cost} segmentation that selects by dynamic programming over a trie of the vocabulary the token sequence minimizing $\sum_i c(t_i)$ with $c(t) = \log_2(\pi_s(t) + 2)$, the code length of a token under Zipf's law given its static rank (the Viterbi tokenizer of Kudo~\cite{kudo2018} with Zipf probabilities in place of learned ones).

\subsection{Information-Theoretic Analysis}

Under Zipf's law with $\alpha \approx 1$, the token of rank $r$ has probability $p_r = 1/((r+1) H_V)$, where $H_V = \sum_{k=1}^{V} 1/k \approx \ln V + \gamma$ and $V = 71{,}161$ distinct tokens occur on enwik8. The 128 one-byte ranks then cover $H_{128}/H_V \approx 5.43/11.75 \approx 46\%$ of all tokens; ranks below 16{,}384 (2 LEB128 bytes) cover another $\approx 41\%$, and the remaining $\approx 13\%$ need 3 bytes. The expected cost is $\bar{L} \approx 0.46 + 0.41 \cdot 2 + 0.13 \cdot 3 = 1.67$ bytes per token.

\begin{figure}[t]
\centering
\includegraphics[width=\columnwidth]{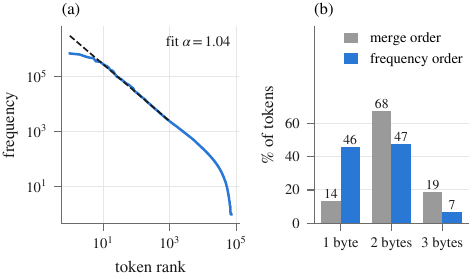}
\caption{(a) BPE token rank vs.\ frequency on enwik8 (cl100k, log-log), with a Zipf fit over the top 1000 ranks ($\alpha = 1.04$). (b) Share of tokens needing 1, 2 or 3 LEB128 bytes with BPE's merge-order IDs and with frequency-ordered IDs.}
\label{fig:zipf}
\end{figure}

In our measurements, BPE's merge-order IDs place only 13.6\% of enwik8 tokens below 128, while with frequency ordering 45.7\% of tokens need 1 byte and 47.4\% need 2 bytes, giving $\bar{L} = 1.61$ bytes and a 41.6\,MB LEB128 stream for 25.8\,M tokens. The prefix code's wider 2-byte band and the minimum-cost segmentation reduce it further to 40.1\,MB. Fig.~\ref{fig:zipf} shows the fitted exponent ($\alpha = 1.04$) and the byte-length shift.

\section{Experiments}

\subsection{Setup}

We evaluate on the standard \textbf{enwik8} benchmark~\cite{mahoney2006} (first 100\,MB of a Wikipedia XML dump) and the larger \textbf{enwik9} (first 1\,GB). We use the \texttt{cl100k\_base} vocabulary from OpenAI~\cite{tiktoken2023} (vocabulary size $\sim$100k) with the minimum-cost segmentation and the automatic rank map. The static rank table is computed from bytes 100--200\,MiB of enwik9, which are disjoint from enwik8, and is used unchanged for every dataset, including the non-English ones. enwik9 contains these bytes, so its results are partly in-sample for the table (10\% of the file).

All transforms run in our C implementation, which links the reference libraries zstd~1.5.5, xz~5.4.5, brotli~1.1.0, zlib~1.3 and bzip2~1.0.8 as backends. We test seven compressor settings: \textbf{zlib-9}~\cite{gailly1995} (LZ77 + Huffman), \textbf{bz2-9}~\cite{seward1996} (BWT + Huffman), \textbf{xz-6} and \textbf{xz-9e}~\cite{pavlov1998} (LZMA2 with 8\,MB and 64\,MB dictionaries; the latter is the strongest LZMA setting), \textbf{zstd-22}~\cite{collet2016} (LZ + finite-state entropy, 128\,MB window), \textbf{brotli-11}~\cite{alakuijala2018} (16\,MB window) and \textbf{PPMd-16}~\cite{shkarin2002} (order-16 PPM, via pyppmd). Ratios are compressed over original size (\%); FOT ratios include the container header and, where used, the hybrid list. Times are wall-clock on one pinned core of an otherwise idle Intel i5-12600K with the input in memory; timing results are medians of three runs (spread below 3\%). Our C implementation is open source at \url{https://github.com/avocardio/fot}.

\subsection{Main Results}

Table~\ref{tab:main} shows the main results: FOT improves every LZ-family compressor.

\begin{table}[t]
\centering
\caption{Compression ratios on enwik8 (100\,MB). Lower is better. FOT ratios include all headers and the 128-token hybrid list (0.3\,KB).}
\label{tab:main}
\begin{tabular}{@{}lccc}
\toprule
Compressor & Raw (\%) & FOT (\%) & Impr.\ (pp) \\
\midrule
zlib-9     & 36.48 & 28.52 & 7.96 \\
bz2-9      & 29.01 & 28.07 & 0.94 \\
xz-6       & 26.38 & 24.14 & 2.24 \\
xz-9e      & 24.83 & 23.46 & 1.37 \\
zstd-22    & 25.33 & 23.89 & 1.44 \\
brotli-11  & 25.74 & 23.78 & 1.96 \\
PPMd-16    & 22.83 & 22.90 & $-0.07$ \\
\bottomrule
\end{tabular}
\end{table}

Simpler compressors that do not model byte statistics beyond order-0 Huffman coding benefit most. PPMd, which builds an order-16 context model with arithmetic coding, is unaffected within noise ($-0.07$\,pp): its model already captures what the frequency ordering exposes, although it runs faster on the shorter stream (18\,s including the transform, vs.\ 30\,s).

\subsection{Ablation: Tokenization vs.\ Reordering}

Table~\ref{tab:ablation} adds one component per row on the full 100\,MB, starting from exact BPE tokens with merge-order IDs.

\begin{table}[t]
\centering
\caption{Ablation on enwik8 (100\,MB), compression ratio (\%). Each row adds one component; all per-file costs included.}
\label{tab:ablation}
\resizebox{\columnwidth}{!}{\begin{tabular}{@{}lcccc}
\toprule
Pipeline & zlib-9 & xz-6 & zstd-22 & brotli-11 \\
\midrule
Raw UTF-8                                & 36.48 & 26.38 & 25.33 & 25.74 \\
BPE tokens, merge-order IDs, LEB128      & 32.68 & 26.01 & 25.49 & 25.29 \\
+ adaptive frequency order (table sent)  & 29.38 & 24.62 & 24.48 & 24.08 \\
+ static table (none sent)               & 29.23 & 24.50 & 24.32 & 23.97 \\
+ prefix code instead of LEB128          & 28.85 & 24.39 & 24.12 & 24.02 \\
+ minimum-cost segmentation              & 28.55 & 24.16 & 23.91 & 23.81 \\
+ automatic list                  & \best{28.52} & \best{24.14} & \best{23.89} & \best{23.78} \\
\bottomrule
\end{tabular}}
\end{table}

Tokenization alone helps zlib by 3.8\,pp and xz and brotli by 0.4\,pp, but \emph{hurts} zstd by 0.16\,pp, because the merge-order IDs produce a stream whose bytes are nearly uniformly distributed within each varint band. Frequency reordering adds 1.0--3.3\,pp, the largest single step for every compressor but zlib. The static table then matches the adaptive order without sending anything (0.1--0.2\,pp, the cost of the table), the prefix code adds 0.1--0.4\,pp for zlib, xz and zstd but costs 0.05\,pp for brotli, and the minimum-cost segmentation adds 0.2--0.3\,pp over exact BPE. The automatic list changes little on in-domain text; it matters on other languages (Section~\ref{sec:cross}).

\subsection{Per-File Table Overhead}
\label{sec:overhead}

Table~\ref{tab:overhead} quantifies why the static table matters. With the adaptive table the method \emph{loses} to raw zstd-22 at 100\,KB and 1\,MB and only breaks even at 10\,MB; with the static table it gains 6.6\,pp at 100\,KB. All numbers in this paper include this overhead.

\begin{table}[t]
\centering
\caption{Adaptive rank-table overhead vs.\ file size (enwik8 prefixes, cl100k). zstd-22 ratios (\%) for raw text, the adaptive table including its cost, and FOT with the static table.}
\label{tab:overhead}
\resizebox{\columnwidth}{!}{\begin{tabular}{@{}lrrrrrr}
\toprule
Size & $U$ & Table & Table (\%) & Raw & Adaptive & Static \\
\midrule
100\,KB & 4{,}830  & 10.8\,KB & 10.80 & 33.75 & 36.71 & \best{27.11} \\
1\,MB   & 21{,}222 & 49.8\,KB & 4.98  & 30.01 & 31.74 & \best{26.52} \\
10\,MB  & 50{,}500 & 121\,KB  & 1.21  & 27.93 & 27.81 & \best{26.08} \\
100\,MB & 71{,}161 & 172\,KB  & 0.17  & 25.33 & 24.48 & \best{23.89} \\
\bottomrule
\end{tabular}}
\end{table}

\subsection{File Size Scaling}

\begin{figure}[t]
\centering
\includegraphics[width=\columnwidth]{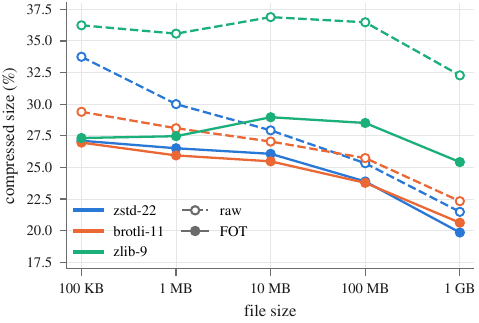}
\vspace{-0.3em}
\caption{Compressed size vs.\ file size (enwik8 prefixes, enwik8, enwik9) for zstd-22, brotli-11 and zlib-9. Solid, filled: FOT; dashed, open: raw.}
\label{fig:scaling}
\vspace{0.5em}
\includegraphics[width=\columnwidth]{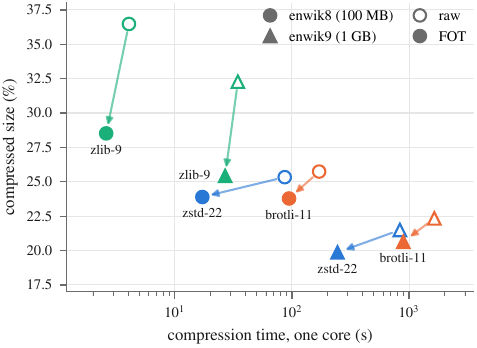}
\vspace{-0.3em}
\caption{Compression time (one core, log scale) vs.\ compressed size for zstd-22, brotli-11 and zlib-9. Circles: enwik8 (medians of three runs); triangles: enwik9 (one run, with an earlier, 20\% slower build of the transform, so its speedups are slightly conservative). Open markers: raw; filled: FOT, including the transform time. For all three compressors, FOT reduces both time and size.}
\label{fig:pareto}
\end{figure}

Fig.~\ref{fig:scaling} shows that gains grow for smaller files (zlib +8.9\,pp and zstd-22 +6.6\,pp at 100\,KB; +6.9 and +1.6\,pp at 1\,GB), where the raw compressor has seen too little data to learn the statistics that the static table supplies for free.

\subsection{Compression Speed}

FOT accelerates compression (Fig.~\ref{fig:pareto}): total time, including the transform, falls from 4.1, 58.8, 86.8 and 170.8\,s to 2.6, 18.2, 17.2 and 94.5\,s for zlib-9, xz-6, zstd-22 and brotli-11 (medians of three runs), a $5.0\times$ speedup for zstd-22 and $1.6$--$3.2\times$ for the others. The transform takes 1.6\,s per 100\,MB on one core (about 0.3\,s on 16 cores, since blocks are independent) and reduces the input from 100\,MB of text to a 40.1\,MB stream that compresses much faster.

Across the 21 settings we tested (levels sampled from zstd 1--22, xz 1--9e, brotli 1--11 and zlib 1--9, plus bz2-9), 16 raw settings are beaten on both time and size by some FOT setting, only one FOT setting (brotli-1) is beaten by a raw one, and every point on the Pareto frontier slower than 1.2\,s is an FOT setting; for example, zstd-3 with FOT (1.8\,s, 29.44\%) matches raw zstd-15 (28.6\,s, 29.44\%). The exceptions are the five fastest raw settings (zstd-1, 3 and 6, brotli-1, zlib-1, each at most 1.2\,s), where the transform costs more than the compressor itself; there a greedy longest-match segmentation (0.7\,s) costs less than half at a 0.5\,pp ratio penalty, and multiple threads reduce it further. FOT does not speed up every level: brotli-9 becomes slower (28\,s vs.\ 26\,s) while compressing 2.4\,pp better.

\subsection{Vocabulary Size}

Table~\ref{tab:vocab} compares three pretrained vocabularies, each with its own static table learned from the same corpus. On English, size matters little: from GPT-2's 50k to o200k's 200k tokens, zlib-9 improves by 0.40\,pp and zstd-22 by only 0.05\,pp. o200k gains 0.9\,pp on Chinese (zstd-22, Table~\ref{tab:cross}) but loses 0.6 and 0.2\,pp on Arabic and Python.

\begin{table}[t]
\centering
\caption{Pretrained vocabularies on enwik8 (100\,MB), compression ratio (\%), each with its own static table.}
\label{tab:vocab}
\begin{tabular}{@{}lccccc}
\toprule
Vocabulary & $|V|$ & zlib-9 & xz-6 & zstd-22 & brotli-11 \\
\midrule
GPT-2     & 50k  & 28.68 & 24.19 & 23.88 & 23.78 \\
cl100k    & 100k & 28.52 & 24.14 & 23.89 & 23.78 \\
o200k     & 200k & \best{28.28} & \best{24.08} & \best{23.83} & \best{23.71} \\
\midrule
\multicolumn{2}{l}{Raw (no preprocessing)} & 36.48 & 26.38 & 25.33 & 25.74 \\
\bottomrule
\end{tabular}
\end{table}

\subsection{Segmentation}

Table~\ref{tab:seg} compares three ways of segmenting text into the cl100k vocabulary. Exact BPE inference is neither the fastest nor the best: greedy longest match is $10\times$ faster than tiktoken and loses 0.25--0.45\,pp, while the minimum-cost segmentation is $4.3\times$ faster and \emph{gains} 0.2--0.3\,pp on every compressor.

\begin{table}[t]
\centering
\caption{Segmentation of enwik8 into the cl100k vocabulary (static table, prefix code). Speed is single-core transform throughput.}
\label{tab:seg}
\resizebox{\columnwidth}{!}{\begin{tabular}{@{}lccccc}
\toprule
Segmentation & MB/s & zlib-9 & xz-6 & zstd-22 & brotli-11 \\
\midrule
Exact BPE (tiktoken) & 14  & 28.84 & 24.39 & 24.11 & 24.02 \\
Greedy longest match & 140 & 29.29 & 24.68 & 24.36 & 24.29 \\
Minimum cost (FOT)   & 60  & \best{28.52} & \best{24.14} & \best{23.89} & \best{23.78} \\
\bottomrule
\end{tabular}}
\end{table}

\subsection{Cross-Dataset Generalization}
\label{sec:cross}

Table~\ref{tab:cross} covers three natural languages and source code, from 98\,KB to 1\,GB, always with the English static table. Every entry is an improvement. On enwik9 the speedups are $3.4\times$ (zstd-22, 835\,s to 245\,s), $2.8\times$ (xz-6) and $1.8\times$ (brotli-11). For Chinese and Arabic the encoder prefixes the table with the file's most frequent tokens (1024 tokens, 2.7\,KB, and 128 tokens, 0.3\,KB).

\begin{table}[t]
\centering
\caption{Cross-dataset generalization (cl100k, English static table). Improvement over raw in pp; all overheads included.}
\label{tab:cross}
\resizebox{\columnwidth}{!}{\begin{tabular}{@{}llcccc}
\toprule
Dataset & Size & \multicolumn{4}{c}{Improvement (pp)} \\
\cmidrule(lr){3-6}
& & zlib-9 & xz-6 & zstd-22 & brotli-11 \\
\midrule
enwik8 (100\,KB subset) & 100\,KB & \best{8.91} & \best{6.14} & \best{6.64} & \best{2.44} \\
enwik8 (10\,MB subset)  & 10\,MB  & 7.91 & 1.70 & 1.85 & 1.57 \\
enwik9 (English)    & 1\,GB    & 6.86 & 1.91 & 1.63 & 1.70 \\
Shakespeare         & 5.4\,MB  & 7.82 & 2.32 & 2.26 & 2.33 \\
Python code         & 98\,KB   & 3.15 & 1.88 & 1.79 & 0.27 \\
Chinese Wikipedia   & 2.0\,MB  & 5.64 & 0.65 & 1.06 & 0.80 \\
\quad with o200k    &          & 8.07 & 1.63 & 1.95 & 1.84 \\
Arabic Wikipedia    & 2.0\,MB  & 4.56 & 1.16 & 1.42 & 0.99 \\
\bottomrule
\end{tabular}}
\end{table}

\subsection{Comparison to Word Replacing Transform}

Table~\ref{tab:wrt} compares our method against a Word Replacing Transform (WRT) baseline~\cite{skibinski2005}, implemented as word-level frequency-ranked replacement with varint encoding and a 0xFF escape byte for word boundaries. It omits the original's capitalization modeling, number encoding and multi-type boundary markers, so the real WRT would likely do somewhat better, making our advantage conservative. FOT outperforms it on every compressor: +7.96\,pp vs.\ +4.74\,pp for zlib and +1.44\,pp vs.\ $-$0.08\,pp for zstd. The subword decomposition handles morphological variants and rare words without a separate entry per surface form, giving fewer distinct entries on enwik8 (71k vs.\ 354k) and a more compact stream (40.1\% vs.\ 67.6\% of the original size).

\begin{table}[t]
\centering
\caption{Comparison to Word Replacing Transform on enwik8 (100\,MB). Ratios include all overheads. WRT was measured with the Python bindings of the same codecs, whose zstd-22 baseline is 25.27\%; its improvement is relative to that baseline.}
\label{tab:wrt}
\begin{tabular}{@{}lccc}
\toprule
Method & zlib-9 (\%) & zstd-22 (\%) & xz-6 (\%) \\
\midrule
Raw              & 36.48 & 25.33 & 26.38 \\
WRT~\cite{skibinski2005}  & 31.74 & 25.35 & 25.51 \\
\textbf{FOT}     & \best{28.52} & \best{23.89} & \best{24.14} \\
\bottomrule
\end{tabular}
\end{table}

\subsection{Comparison to State of the Art}

Table~\ref{tab:sota} places our results in the context of the Large Text Compression Benchmark~\cite{mahoney2006}. With xz tuned like the benchmark's xz entry (preset 9e, lc=4, pb=0, dictionary $\geq$ input), FOT compresses enwik9 to 189,382,412 bytes (189,923,156 with our decompressor, vocabulary and table), below every LZ-based entry (best: plzma, 193,240,160; tuned xz alone: 197,331,816). On enwik8 the best LZ entries stay ahead (comprolz 22.81\,MB vs.\ 23.42\,MB for FOT with tuned xz); they already use a per-file word dictionary, which FOT nearly matches when replacing it (in our build, comprolz on the FOT stream 23.15\,MB vs.\ 22.88\,MB with its own dictionary). The enwik9 result is partly in-sample, since the static table was learned from 10\% of enwik9; the benchmark permits this because the table is counted in the decompressor size.

\begin{table}[t]
\centering
\caption{Comparison on enwik8 (100\,MB). SOTA results from the Large Text Compression Benchmark~\cite{mahoney2006}. FOT applies to the practical LZ-family compressors below the line.}
\label{tab:sota}
\begin{tabular}{@{}lcc}
\toprule
Method & Ratio (\%) & Category \\
\midrule
cmix v21~\cite{knoll2024}    & 14.62 & Neural/context mixing \\
NNCP v3.2~\cite{bellard2021} & 14.92 & Transformer \\
PAQ8px     & 15.85 & Context mixing \\
PPMd-16                       & 22.83 & Statistical \\
\midrule
\textbf{FOT + xz-9e}        & \best{23.46} & LZ + FOT \\
\textbf{FOT + brotli-11}    & \best{23.78} & LZ + FOT \\
\textbf{FOT + zstd-22}      & \best{23.89} & LZ + FOT \\
xz-9e                         & 24.83 & LZ baseline \\
zstd-22                       & 25.33 & LZ baseline \\
brotli-11                     & 25.74 & LZ baseline \\
\textbf{FOT + zlib-9}       & \best{28.52} & LZ + FOT \\
zlib-9                        & 36.48 & LZ baseline \\
\bottomrule
\end{tabular}
\end{table}

\section{Discussion}

\textbf{Why zlib benefits most.} zlib has a 32\,KB window and per-block Huffman codes with no context modeling of literals. The transformed stream has about half of its tokens as single bytes below 128, lowering literal entropy, and each stream byte carries 2.5 bytes of text, so the window and the 258-byte maximum match cover $2.5\times$ more text. zstd's entropy coders already adapt to skewed bytes, and LZMA and brotli model literals in context with much larger windows, so they capture part of this structure themselves.

\textbf{Why bz2 and PPMd benefit less.} bz2 applies the Burrows-Wheeler Transform (BWT), which already groups similar contexts and assigns small integers via move-to-front coding, a conceptually similar reordering to FOT. BPE tokenization disrupts BWT's byte-level context grouping, so the net gain is only 0.94\,pp. PPMd~\cite{shkarin2002} builds a context-conditioned probability model and uses arithmetic coding, which already approaches the order-$k$ entropy, and is unaffected ($-$0.07\,pp). The same holds for grammar-based compression: GLZA already builds a file-specific phrase dictionary, and on a 10\,MB enwik8 sample FOT makes it 1.87\,pp worse (24.60\% vs.\ 22.73\%), although $2.4\times$ faster. FOT is thus \emph{complementary} to, not a substitute for, models that already capture frequency or context structure.

\textbf{The vocabulary is the asset, not the tokenizer.} Exact BPE inference reproduces the training-time segmentation that a language model expects, but a compressor has no such constraint. The minimum-cost segmentation reuses the vocabulary and picks, among all valid segmentations, the one with the shortest estimated code, which is both better and faster than BPE inference (Table~\ref{tab:seg}).

\textbf{Why FOT is faster.} The stream is $\sim$40\% of the text's size and its concentrated small values simplify match finding, so the compressor's savings exceed the transform's cost. This speed--ratio improvement was not reported in concurrent work~\cite{erdogan2026}.

\textbf{Limitations.} (1) \emph{Static table domain}: the shipped table is learned from English Wikipedia (400\,KB once, not per file). Chinese and Arabic text relies on the hybrid list (0.3--2.7\,KB per file) and gains less than a matching table would (Table~\ref{tab:cross}). (2) \emph{Decompression speed}: the inverse transform takes 0.14\,s per 100\,MB on one core (about 700\,MB/s; 0.03\,s on 16 cores), small next to xz (1.02 to 1.16\,s in total) but noticeable for zstd (0.15 to 0.22\,s) and brotli (0.27 to 0.43\,s); the method remains better suited to storage and archival than to latency-critical decoding.
 (3) \emph{Domain}: binary, compressed or encrypted data lacks Zipfian token statistics and will not benefit. (4) \emph{Streaming}: with the static table the transform is single-pass; the hybrid list requires a first pass over the data.

\textbf{Future work.} Promising directions include compression-aware tokenizers (cf.\ LZ-aware BPE in~\cite{erdogan2026}), tuning compressor parameters to the FOT stream's statistics, a small static dictionary for the transformed stream in the manner of brotli, and extension to other data with Zipfian statistics. Finally, a custom BPE vocabulary trained on the target data, such as logs or source code, could compress it further, at the cost of shipping a dedicated vocabulary.

\section{Conclusion}

We presented frequency-ordered tokenization, a simple, theoretically grounded preprocessing step that improves lossless text compression by 1.3--8 percentage points across LZ-family compressors, with negligible per-file overhead. It exploits Zipf's law through frequency-ranked IDs and a byte-oriented prefix code and, freed from the tokenizers' inference rule, chooses the segmentation with the shortest code. For all but the fastest compressor settings, FOT improves the speed--ratio tradeoff, a Pareto improvement over the raw compressors.

Its vocabularies come already trained for large language models and are used frozen, so FOT needs no tokenizer training and no change to existing compressors; it is available as a drop-in C transform. As the volume of natural language text generated and stored continues to accelerate, driven by web-scale corpora, conversational AI systems, and the data pipelines surrounding large language models, even modest improvements in compression ratio compound into substantial savings at scale.

\balance

\end{document}